\documentclass[aps,pre,showpacs,superscriptaddress,12pt]{revtex4}

\usepackage{amsmath}
\usepackage{graphicx}

\begin{document}

\title{Modulational instability of spatially broadband nonlinear optical pulses
  in four-state atomic systems}
  
\author{M. Marklund}
\altaffiliation[Also at: ]{Centre for Fundamental Physics, Rutherford Appleton Laboratory,
  Chilton, Didcot, Oxon OX11 OQX, U.K.}
\affiliation{Centre for Nonlinear Physics, Department of Physics, 
  Ume{\aa} University, SE--901 87 Ume{\aa}, Sweden}

\author{P. K. Shukla}
\altaffiliation[Also at: ]{Centre for Nonlinear Physics, Department of Physics, 
  Ume{\aa} University, SE--901 87 Ume{\aa}, Sweden}
\altaffiliation{Centre for Fundamental Physics, Rutherford Appleton Laboratory,
  Chilton, Didcot, Oxon OX11 OQX, U.K.}
\altaffiliation{Department of Physics, University of Strathclyde, Glasgow, Scotland, 
  G4 ONG, UK}
\altaffiliation{Centro de F\'{i}sica dos Plasmas, Instituto Superior T\'{e}cnico, 1049-001 Lisboa, 
  Portugal}
\affiliation{Institut f\"ur Theoretische Physik IV and Centre for Plasma Science 
  and Astrophysics, Fakult\"at f\"ur Physik und Astronomie, Ruhr-Universit\"at Bochum, 
  D--44780 Bochum, Germany}

\author{R. Bingham}
\altaffiliation[Also at: ]{Centre for Fundamental Physics, Rutherford Appleton Laboratory,
  Chilton, Didcot, Oxon OX11 OQX, U.K.}
\altaffiliation{Department of Physics, University of Strathclyde, Glasgow, Scotland, 
  G4 ONG, UK}
\affiliation{Space Science \& Technology Department, Rutherford Appleton Laboratory,
  Chilton, Didcot, Oxon OX11 OQX, U.K.}

\author{J. T. Mendon\c{c}a}
\altaffiliation[Also at: ]{Centre for Fundamental Physics, Rutherford Appleton Laboratory,
  Chilton, Didcot, Oxon OX11 OQX, U.K.} 
\affiliation{Centro de F\'{i}sica dos Plasmas, Instituto Superior T\'{e}cnico, 1049-001 Lisboa, 
  Portugal}

\begin{abstract}
  The modulational instability of broadband optical pulses in a 
 four-state atomic system is investigated. In particular,
  starting from a recently derived generalized nonlinear 
  Schr\"odinger equation, a wave-kinetic equation is derived.
  A comparison between coherent and random phase
  wave states is made. It is found that the spatial spectral
  broadening can contribute to the nonlinear stability of ultra-short optical 
  pulses. In practical terms, this could be achieved by using
  random phase plate techniques. 
\end{abstract}

\maketitle

\newpage

The propagation of ultra-short optical pulses in nonlinear media is of great importance in
a wide variety of applications, e.g. fibre optical systems \cite{shukla-rasmussen,shukla-marklund}, 
atmospheric remote sensing using femto-second laser pulses in air \cite{braun-etal,berge-etal,marklund-shukla}, 
and inertial confinement fusion \cite{koenig-etal}. The nonlinear propagation of ultra-short 
optical pulses can be modeled by using the nonlinear Schr\"odinger equation with 
short-pulse nonlinear derivative corrections, which can give rise to the filamentation
of light pulses, and subsequent formation of light pipes. Modulational and filamentational 
instabilities therefore sometimes pose problems concerning pulse propagation in nonlinear media.
Thus, for propagation times 
much longer than the typical pulse length, it is of importance to find means
for the nonlinear stabilization of such optical pulses. 

In this Brief Report, we investigate the statistical properties of a generalized
nonlinear Schr\"odinger equation, taking into account the Kerr nonlinearity,
linear absorption, nonlinear dispersion, delay in nonlinear refractive index, 
third-order dispersion, differential absorption, and diffraction. The equation 
is of relevance for Raman excited four-state atomic systems. We derive an equivalent 
wave-kinetic equation that governs the propagation of optical quasi-particles, 
enabling us to study the effect of partial pulse coherence. The newly derived 
equation is analyzed for the case of spectral pulse broadening, and it is found
that the relevant growth rate can be significantly reduced if an appropriate random 
phase techniques are used.

\begin{figure}
\includegraphics[width=0.85\textwidth]{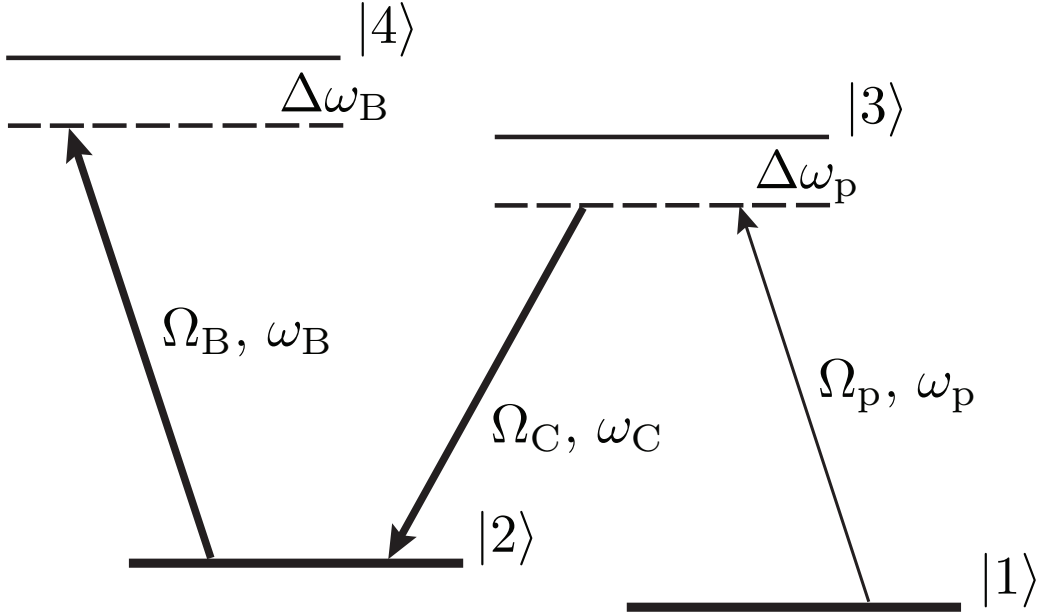}
\caption{Energy-level diagram showing the excitation of the 
  relevant four-state atomic system. Here $\omega_{\mathrm{p}}$
  is the center frequency of the weak probe field, while $\omega_{\mathrm{B}}$
  and $\omega_{\mathrm{C}}$ are the frequencies of two strong cw control fields. 
  The one-photon detuning frequencies are defined according to $\Delta\omega_{\mathrm{p}} 
  = \omega_{31} - \omega_{\mathrm{p}}$ and $\Delta\omega_{\mathrm{B}} 
  = \omega_{42} - \omega_{\mathrm{B}}$.
  Moreover, $2\Omega_{\mathrm{p}}$, $2\Omega_{\mathrm{B}}$, and $2\Omega_{\mathrm{C}}$
  are the Rabi frequencies for the indicated transitions, and two-photon
  resonance is assumed to always be maintained \cite{hang-etal,wu-deng}.}
\end{figure}

Recently, Hang {\it et al.} \cite{hang-etal} derived a generalized nonlinear Schr\"odinger
equation (NLSE) which  describes the nonlinear propagation of optical pulses in Raman 
excited four-state atomic systems (see Fig.\ 1). In a dimensionless form,  the NLSE can be written as 
\begin{equation}\label{eq:nlse}
  i\partial_zu + \partial_t^2u + 2|u|^2u = 
  - i\left[ 
    d_0u - d_1\partial_t(|u|^2u) - d_2u\partial_t|u|^2 - d_3\partial_t^3u
  \right]
  + d_4\partial_tu - d_5\nabla_{\perp}^2u ,
\end{equation}
where $u$ is the normalized pulse amplitude and $d_j$ for $j = 0$ 
to $5$ is the dimensionless coefficients for  linear absorption,
 nonlinear dispersion, delay in nonlinear refractive index, third-order
dispersion, differential absorption, and diffraction, respectively.
Equation (\ref{eq:nlse}) thus describes the nonlinear propagation of ultra-short
optical pulses when the terms in the right-hand side of (1) are significant (i.e.\
$d_j \sim 1$) \footnote{However, we note that the terms proportional to
$d_j$, $j = 1, 2, 3$, on the right hand side of
Eq.\ (\ref{eq:nlse}) is still part of a perturbative expansion, and should be treated as small
compared to the terms on the left hand side. Thus, the time variations in the 
amplitude $u$ must be weak.}.
In what follows, we will look at the one-dimensional problem, and thus 
set $d_5 = 0$.

In order to analyze the spectral evolution of partially coherent optical 
pulses, we next introduce the Fourier transform of the two-point
correlation function, i.e.\ the Wigner function \cite{wigner,moyal,mendonca}
\begin{equation}\label{eq:wigner}
  \rho(z,t,\omega) = \frac{1}{2\pi}\int d\tau\, e^{-i\omega\tau}
    \langle u^*(z,t + \tau/2)u(z,t - \tau/2) \rangle ,
\end{equation}
where the angular bracket denotes the ensemble average 
\cite{klimontovich}. The Wigner function represents
a generalized distribution function for optical
quasi-particles. The Wigner method \cite{anderson-etal}, as well as 
the equivalent mutual coherence method \cite{demetrios-etal},
have been used to analyze the modulational instability in the ``standard"
nonlinear Schr\"odinger picture, in which the right-hand side of 
Eq.\ (\ref{eq:nlse}) is identically zero. From the definition (\ref{eq:wigner})
follows the normalization
\begin{equation}
  I(z,t) \equiv \langle|u(z,t)|^2\rangle = \int d\omega\,\rho(z,t,\omega) , 
\end{equation}
giving the pulse intensity in terms of the Wigner function.

Applying the $z$-derivative to the definition (\ref{eq:wigner}), 
we obtain the wave-kinetic equation for optical quasi-particles
\begin{eqnarray}
  &&\!\!\!\!\!\!\!
  \partial_z\rho - 2\omega\partial_t\rho 
  - 4I\sin\left( \tfrac{1}{2}\stackrel{\leftarrow}{\partial}_t%
    \stackrel{\rightarrow}{\partial}_{\omega}\right)\rho
  = -2d_0\rho - d_1\left[ 
    2\omega I\sin\left( \tfrac{1}{2}\stackrel{\leftarrow}{\partial}_t%
    \stackrel{\rightarrow}{\partial}_{\omega}\right)\rho
    - \partial_t\left( I
      \cos\left( \tfrac{1}{2}\stackrel{\leftarrow}{\partial}_t%
    \stackrel{\rightarrow}{\partial}_{\omega}\right)\rho
    \right)
  \right]
\nonumber \\ && \quad
  + 2d_2(\partial_t I)\cos\left( \tfrac{1}{2}\stackrel{\leftarrow}{\partial}_t%
    \stackrel{\rightarrow}{\partial}_{\omega}\right)\rho
  - d_3\left( 3\omega^2 -\tfrac{1}{4}\partial_t^2\right)\partial_t\rho - 2d_4\omega\rho .
\label{eq:kinetic}
\end{eqnarray}
The wave-kinetic equation (\ref{eq:kinetic}) determines the phase space evolution
of partially coherent optical wavepackets. In the low-frequency limit, we can retain 
only the first term in the operator expansions to obtain
\begin{eqnarray}
  \partial_z\rho - 2\omega\partial_t\rho 
  - 2\partial_tI\,\partial_{\omega}\rho
  = -2d_0\rho - d_1\left[ 
    \omega \partial_tI\,\partial_{\omega}\rho
    - \partial_t( I\rho )
  \right]
  + 2d_2\rho\partial_tI
  - 3d_3\omega^2\partial_t\rho - 2d_4\omega\rho .
\label{eq:kinetic-low}
\end{eqnarray}
Here we see that the terms in the left-hand side of (\ref{eq:kinetic-low})
resemble the terms in a Vlasov equation.  Moreover, Eqs.\ (\ref{eq:kinetic}) 
and (\ref{eq:kinetic-low}) have a time-independent solution
\begin{equation}
  \bar{\rho}(z,\omega) = \rho_0(\omega)e^{-2(d_0 + d_4\omega)z} ,
\end{equation}
exhibiting a spatial diffusive influence of the linear and differential absorption.
For a coherent distribution, i.e.\ $\rho_0(\omega) = I_0\delta(\omega - \omega_0)$
for some frequency $\omega_0$, we obtain the intensity $\bar{I}(z) = 
I_0\exp[-2(d_0 + d_4\omega_0)z]$, while for a Gaussian distribution 
$\rho_0(\omega) = (I_0/\sqrt{2\pi}\,\Delta)\exp[-(\omega - \omega_0)^2/2\Delta^2]$
with the spectral width $\Delta$, we obtain the intensity $\bar{I}(z) = 
I_0\exp[-2(d_0 + d_4\omega_0 - d_4\Delta^2z)z]$. We note that the finite
spectral width in the latter case gives rise to \emph{growing} behavior 
for after a certain distance of propagation. In fact, the spatial intensity 
decay halts for $z_{\mathrm{crit}} = (d_0 + d_4\omega_0)/d_4\Delta^2$,
indicating the breakdown of the above model, as we do not expect growing 
modes from loss terms. For a broad spectral distribution, this distance may
be short.

Next, we analyze the modulational instability of broadband optical pulses that are 
governed by Eq.\ (\ref{eq:kinetic}). Letting $\rho(z,t,\omega) = \bar{\rho}(z,\omega) 
+ \rho_1(z,\omega)\exp(iKz - i\Omega t)$ in the latter, where $|\rho_1| \ll \bar{\rho}$, we
linearize the wave kinetic equation against $\rho_1$. If the dimensionless perturbation 
wavenumber satisfies $K \gg d_0, d_4\omega$, the perturbation wavelength is 
much smaller than the decay length. Thus, we may treat the background distribution as a
constant, and neglect the terms containing $d_0$ and $d_4\omega$.
We then obtain the nonlinear dispersion relation 
\begin{equation}\label{eq:disprel}
  1 = \int d\omega\frac{\left[2 - d_1\left(\omega + \Omega/2 \right) 
      - d_2\Omega\right]\bar{\rho}(z,\omega - \Omega/2) 
    - \left[2 - d_1\left(\omega + \Omega/2 \right) 
      + d_2\Omega\right]\bar{\rho}(z,\omega + \Omega/2)}%
    {K + 2\omega\Omega + d_1\Omega\bar{I} - d_3\left(3\omega^2 + \Omega^2/4\right)\Omega} .
\end{equation}
The case of higher order dispersive corrections to the nonlinear Schr\"odinger equation in 
the Wigner picture has been treated in the wavenumber domain in Ref.\ \cite{lukas}.

In the case of coherent optical pulses, we let $\rho_0(\omega) = I_0\delta(\omega - \omega_0)$. 
Here the dispersion relation (\ref{eq:disprel}) reduces to
\begin{eqnarray}
&&
  K = -\left[2\omega_0 + (2d_1 + d_2)\bar{I} - d_3(3\omega_0^2 + \Omega^2)\right]\Omega
\nonumber \\ &&\qquad\,\,
   \pm \left[ 
     (d_1 + d_2)^2\bar{I}^2\Omega^2 + (1 - 3d_3\omega_0)^2\Omega^4  
     - 2(2 - d_1\omega_0)(1 - 3d_3\omega_0)\bar{I}\Omega^2 \right]^{1/2} .
\end{eqnarray}
For simplicity, we focus on the case $\omega_0 = 0$, at which we have the growth rate
$\Gamma = -\mathrm{Im}(K)$
\begin{equation}\label{eq:growthrate-coherent}
  \Gamma = \left[ 4\bar{I}\Omega^2 - \Omega^4
     - (d_1 + d_2)^2\bar{I}^2\Omega^2
   \right]^{1/2} ,
\end{equation}
where the nonlinear dispersion and delay in the nonlinear refractive index gives
a stabilizing effect to the regular Kerr modulational instability (see Fig.\ 2). We note that 
the modulational instability follows as a consequence of the nonlinear pulse propagation
equation (1) 
governed by the underlying dyamics of the four-state atomic system \cite{wu-deng}.
The latter system is in the form of a four-wave mixing set of equations, and it is
therefore perhaps not surprising that the modulational instability occurs for the 
corresponding nonlinear pulse propagation.  
However, other level configurations may in principle posses similar instability
properties, but the issue of the uniqueness of such instabilities to the nonlinear 
four-state atomic system is left for future research. 

If the background pulse $\Psi_0$ has a random phase, with a coherence
width $\Delta$, this corresponds to a Lorentz distribution
\begin{equation}\label{eq:lorentz}
  \rho_0(\omega) = \frac{I_0}{\pi}\,\frac{\Delta}{\omega^2 + \Delta^2} .
\end{equation}
For the case of $d_4 = 0$, i.e.\ vanishing differential absorption, we 
can integrate the dispersion relation (\ref{eq:disprel}) for $\rho_0$
given by the Lorentz distribution (\ref{eq:lorentz}) to obtain
\begin{eqnarray}
&&
  K = -\left[(2d_1 + d_2)\bar{I} - d_3\Omega^2\right]\Omega 
  - 3d_3\Delta^2\Omega + 2i\Delta\Omega
\nonumber \\ &&\qquad \,\,
  \pm \left[
    (d_1 + d_2)^2\bar{I}^2\Omega^2 + (1 + 3id_3\Delta)^2\Omega^4  
     - 2(2 + id_1\Delta)(1 + 3id_3\Delta)\bar{I}\Omega^2
  \right]^{1/2} .
\label{eq:disprel-lorentz}
\end{eqnarray}
From the dispersion relation (\ref{eq:disprel-lorentz}) we see that 
the spectral broadening interact in nontrivial ways with the modifications
of the nonlinear Sch\"odinger equation. For a pure Kerr nonlinearity, 
i.e.\ $d_j = 0$ for all $j$, the random phase of the background pulse 
will give rise to a reduction of the modulational instability growth 
rate \cite{anderson-etal,demetrios-etal}.  Here we find that the random 
phase will also contribute to the real part of $K$, and the imaginary 
contribution via interactions between the corrections to the Kerr nonlinearity 
and spectral broadening. As noted in \cite{lukas}, higher order dispersive 
effects may affect the modulational instability growth rate, but on the perturbative level, third order dispersion only gives a shift in the real wavenumber. This can be seen by setting 
$d_1 = d_2 = 0$, and $d_3\Delta, d_3\Omega \ll 1$. Linearizing Eq.\ (\ref{eq:disprel-lorentz})
we obtain
\begin{equation}
  K = d_3\Omega^3
  - 3d_3\Delta^2\Omega + 2i\Delta\Omega
  \pm i(4\bar{I}\Omega^2 - \Omega^4)^{1/2} 
  \pm \frac{3d_3\Delta\Omega^2}{2(4\bar{I}\Omega^2 - \Omega^4)^{1/2}}\left(
    2\Omega^2 - 4\bar{I}   
  \right) ,
  \label{eq:pert}
\end{equation}
and we see that the third order dispersion only contributes to the real wavenumber shift.

We have solved for the modulational instability growth rate numerically and plotted
the result for different parameter values in Figs.\ 2 and 3. We note that
when $\Delta = 0$ we regain the coherent growth rate 
(\ref{eq:growthrate-coherent}). Moreover, the third order
dispersion only couples to $\Delta$ and is thus not present
in the coherent case. 

From Fig.\ 3 it is clear that for $d_3 = 0$, the partial coherence
of the background pulse gives rise to a reduced growth rate, and
can thus act as a means for stabilizing optical pulses in four-state
atomic systems described by Eq.\ (\ref{eq:nlse}). 
The effect of 
third-order dispersion is to further reduce the instability growth rate, as expected, and the 
new branches in the modulational
instability growth that occurs for high frequency  
perturbations, i.e.\ $\Omega \geq 1$, are not valid, as the assumptions underlying
the derivation of Eq.\ (\ref{eq:nlse}) are no longer satisfied. Thus, the strongly growing modes
depicted in Fig.\ 3 for $\Omega \geq 1$ are not physical.

\begin{figure}
  \includegraphics[width=0.9\columnwidth]{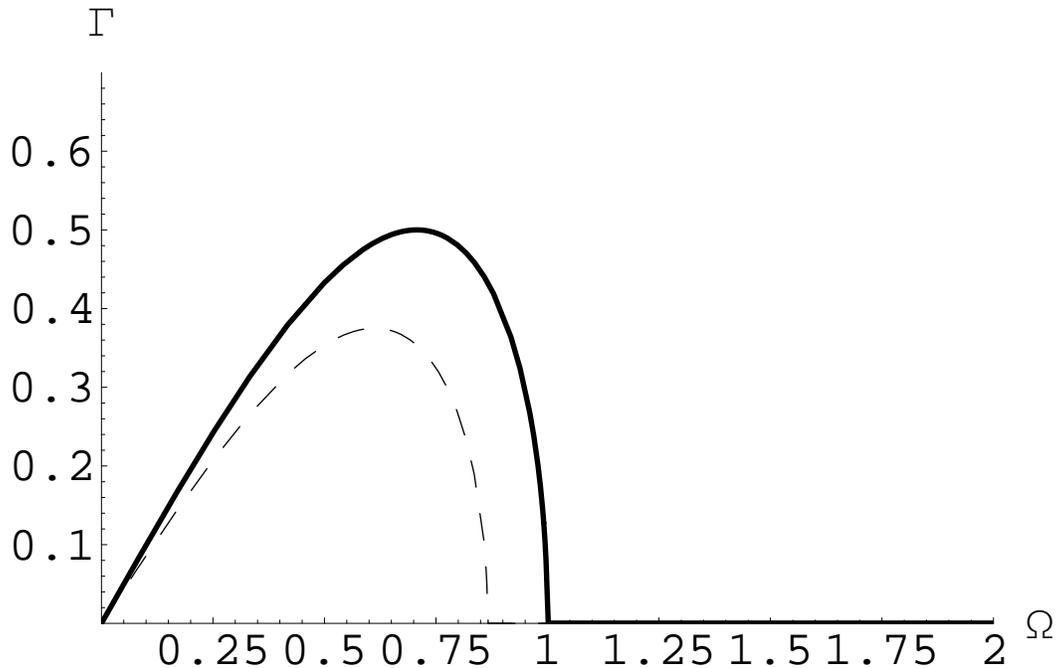}
  \caption{The coherent modulational instability growth rate 
  $\Gamma$ plotted as a function of $\Omega$, using Eq.\ (\ref{eq:disprel-lorentz})
  with $\Delta = 0$ (see Eq.\ (\ref{eq:growthrate-coherent})). The full thick curve has 
  $d_1 = d_2 = d_3 = d_4 = 0$, giving the Kerr modulational instability growth rate, 
  while the dashed curve has $d_1 = d_2 = 1$ and $d_3 = d_4 = 0$. We see
  that the effect of nonlinear dispersion and delay in nonlinear refractive index
  is to reduce the modulational instability growth rate.}
\end{figure}

\begin{figure}
  \includegraphics[width=0.9\columnwidth]{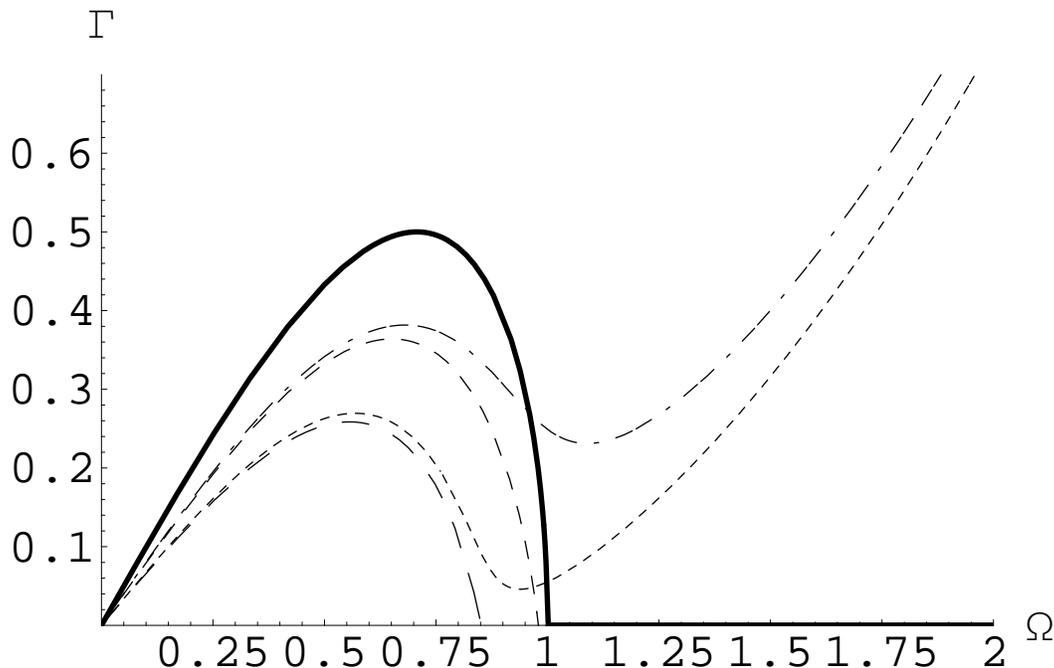}
  \caption{The incoherent modulational instability growth rate 
  $\Gamma$ plotted as a function of $\Omega$, using Eq.\
  (\ref{eq:disprel-lorentz}), for different parameter combinations.
  The full thick curve is the coherent Kerr modulational instability
  growth rate, i.e.\ $d_1 = d_2 = d_3 = d_4 = \Delta =0$. 
  The remaining curves has $\Delta = 0.1$ and $d_4  =0$. From the top 
  the dashed-dotted curve has $d_1 = d_2 = 0$
  and $d_3 = 1$, the dashed curve has $d_1 = d_2 = d_3 = d_4 = 0$,
  the dotted curve has $d_1 = d_2 = d_3 =1$, and the curve with long dashes
  has $d_1 = d_2 = 1$ and $d_3 = 0$. The strongly growing modes for 
  $\Omega \geq 1$ are however not physical, as the underlying assumptions 
  of Eq.\ (\ref{eq:nlse}) are no longer satisfied [see Eq.\ (\ref{eq:pert})].}
\end{figure}

In conclusion, we have presented an investigation of the modulational 
instability of broadband optical pulses in Raman excited four-state 
atomic systems. For this purpose we have obtained a wave-kinetic equation 
from the modified NLSE, which accounts for numerous non-ideal effects
embedded in the right-hand side of (1). Using standard technique, we
then derive a nonlinear dispersion relation from the wave kinetic equation.
The nonlinear dispersion is analyzed for coherent and broadband spectra 
of optical pulses. It is found that the growth rate of the modulational 
instability is reduced in the presence of a Lorentzian optical pulse
distribution. A reduced growth rate insures stability of optical 
pulses over long distances in four-state atomic systems.

\acknowledgments
This research was partially supported by the Swedish Research Council, as
well as by Centre for Fundamental Physics at the Rutherford Appleton
Laboratory, Chilton, Didcot, United Kingdom.


\newpage

\begin{thebibliography}{99}

  \bibitem{shukla-rasmussen}
  P. K. Shukla and J. Juul Rasmussen, Opt. Lett. \textbf{11}, 171 (1986). 
  
  \bibitem{shukla-marklund}
  P. K. Shukla and M. Marklund, Opt. Lett. \textbf{30}, 2548 (2005).
  
  \bibitem{braun-etal}
  A. Braun, G. Korn, X. Liu, D. Du, J. Squier, and G. Mourou, Opt. Lett. \textbf{20}, 73 (1995). 
  
  \bibitem{berge-etal}
  L. Berg\'e, S. Skupin, F. Lederer, G. M\'ejean, J. Yu, J. Kasparian, E. Salmon, J. P. Wolf, 
  M. Rodriguez, L. W\"oste, R. Bourayou, and R. Sauerbrey, Phys. Rev. Lett. \textbf{92}, 
  225002 (2004). 
  
  \bibitem{marklund-shukla}
  M. Marklund and P. K. Shukla, Opt. Lett. \textbf{31}, 1884 (2006).
  
  \bibitem{koenig-etal}
  M. Koenig, B. Faral, J. M. Boudenne, D. Batani, A. Benuzzi, and
  S. Bossi, Phys. Rev. E \textbf{50}, R3314 (1994). 
  
  \bibitem{hang-etal}
  C. Hang, G. Huang, and L. Deng, Phys. Rev. E \textbf{73}, 036607 (2006).
  
  \bibitem{wu-deng}
  Y. Wu and L. Deng, Phys. Rev. Lett. \textbf{93}, 143904 (2004).
  
   \bibitem{wigner}
  E. P. Wigner, Phys. Rev. \textbf{40}, 749 (1932); 
  
  \bibitem{moyal}
  J. E. Moyal, Proc. Cambridge Philos. Soc. \textbf{45}, 99 (1949). 
  
  \bibitem{mendonca}
  J. T. Mendon\c{c}a, \textit{Theory of Photon Acceleration} (IOP Publishing, Bristol, 2001).
  
  \bibitem{klimontovich}
  Yu. L. Klimontovich, \textit{The Statistical Theory of Non-Equilibrium Processes in a Plasma} 
  (Pergamon Press, Oxford, 1967). 
  
   \bibitem{anderson-etal}
  D. Anderson, B. Hall, M. Lisak and M. Marklund, Phys. Rev. E {\bf 65}, 046417 (2002);
  B. Hall, M. Lisak, D. Anderson, R. Fedele, and V. E. Semenov, {\it ibid.} {\bf 65}, 035602 (2002);
  M. Marklund and P. K. Shukla, Rev. Mod. Phys. {\bf 78}, No. 2, in press (2006).
  
  \bibitem{demetrios-etal}
  M. Soljacic, M. Segev, T. Coskun, D. N. Christodoulides, and A. Vishwanath,
  Phys. Rev. Lett. \textbf{84}, 467 (2000).
  
  \bibitem{lukas}
  L.\ Helczynski, M. Lisak, and D. Anderson, Phys. Rev. E \textbf{67}, 026602 (2003).

\end{thebibliography}
\end{document}